\documentclass[acmsmall,screen]{acmart}

\usepackage{subfig}
\usepackage{adjustbox} %
\usepackage{ragged2e}
\usepackage{algorithm}
\usepackage{listings}
\usepackage{enumitem}
\usepackage{tikz}
\usepackage{xcolor}
\usepackage{verbatim}
\usepackage{multirow}
\usepackage{graphicx}
\usepackage{wrapfig}
\usepackage{tcolorbox}
\usepackage{tabularx}
\usepackage{hyperref}
\usepackage{amsmath}
\usepackage{booktabs}
\usepackage{amsfonts}
\usepackage{algorithmic}
\usepackage{array}
\usepackage{textcomp}
\usepackage{stfloats}
\usepackage{url}

\definecolor{githubbgcolor}{HTML}{F6F8FA}
\definecolor{githubcomment}{HTML}{6A737D}
\definecolor{githubkeywords}{HTML}{D73A49}
\definecolor{githubstrings}{HTML}{032F62}
\definecolor{githubgreen}{HTML}{22863A}
\definecolor{githubred}{HTML}{D73A49}
\definecolor{githubblack}{HTML}{24292E}
\definecolor{githubblue}{HTML}{005CC5} %

\AtBeginDocument{%
  \providecommand\BibTeX{{%
    \normalfont B\kern-0.5em{\scshape i\kern-0.25em b}\kern-0.8em\TeX}}}

\setcopyright{acmcopyright}
\copyrightyear{2025}
\acmYear{2025}
\acmDOI{XXXXXXX.XXXXXXX}

\acmJournal{TOSEM}
\acmVolume{0}
\acmNumber{0}
\acmArticle{1}
\acmMonth{0}

\begin{document}

\title{TestART: Improving LLM-based Unit \underline{Test}ing via Co-evolution of \underline{A}utomated Generation and \underline{R}epair I\underline{t}eration}

\author{Siqi Gu}
\email{siqi.gu@smail.nju.edu.cn}
\orcid{0000-0001-5514-6734}
\affiliation{
  \institution{State Key Laboratory for Novel Software Technology, Nanjing University}
  \city{Nanjing}
  \state{Jiangsu}
  \country{China}
}

\author{Quanjun Zhang}
\email{quanjun.zhang@smail.nju.edu.cn}
\orcid{0000-0002-2495-3805}
\affiliation{
  \institution{State Key Laboratory for Novel Software Technology, Nanjing University}
  \city{Nanjing}
  \state{Jiangsu}
  \country{China}
}

\author{Kecheng Li}
\email{522024320081@smail.nju.edu.cn}
\orcid{0009-0008-3084-4921}
\affiliation{
  \institution{State Key Laboratory for Novel Software Technology, Nanjing University}
  \city{Nanjing}
  \state{Jiangsu}
  \country{China}
}

\author{Chunrong Fang}
\email{fangchunrong@nju.edu.cn}
\authornotemark[1]
\orcid{0000-0002-9930-7111}
\affiliation{
  \institution{State Key Laboratory for Novel Software Technology, Nanjing University}
  \city{Nanjing}
  \state{Jiangsu}
  \country{China}
}

\author{Fangyuan Tian}
\orcid{}
\affiliation{%
 \institution{The State Key Laboratory for Novel Software Technology, Nanjing University}
 \city{Nanjing}
 \state{Jiangsu}
 \country{China}}

\author{Liuchuan Zhu}
\email{zhuliuchuan1@huawei.com}
\orcid{}
\affiliation{%
 \institution{Huawei Cloud Computing Technologies Co., Ltd.}
 \state{Beijing}
 \country{China}}

\author{Jianyi Zhou}
\email{zhuliuchuan1@huawei.com}
\orcid{0000-0002-4867-5416}
\affiliation{%
 \institution{Huawei Cloud Computing Technologies Co., Ltd.}
 \state{Beijing}
 \country{China}}

\author{Zhenyu Chen}
\email{zychen@nju.edu.cn}
\orcid{0000-0002-9592-7022}
\affiliation{
  \institution{State Key Laboratory for Novel Software Technology, Nanjing University}
  \city{Nanjing}
  \state{Jiangsu}
  \country{China}
}

\begin{abstract}
Unit testing is crucial for detecting bugs in individual program units but consumes time and effort. Recently, large language models (LLMs) have demonstrated remarkable capabilities in generating unit test cases. However, several problems limit their ability to generate high-quality unit test cases: (1) compilation and runtime errors caused by the hallucination of LLMs; (2) lack of testing and coverage feedback information restricting the increase of code coverage; and (3) the repetitive suppression problem causing invalid LLM-based repair and generation attempts. 
To address these limitations, we propose \textbf{TestART}, a novel unit test generation method. TestART improves LLM-based unit testing via co-evolution of automated generation and repair iteration, representing a significant advancement in automated unit test generation. 
TestART leverages the template-based repair strategy to effectively fix bugs in LLM-generated test cases for the first time. Meanwhile, TestART extracts coverage information from successful test cases and uses it as coverage-guided testing feedback. It also incorporates positive prompt injection to prevent repetition suppression, thereby enhancing the sufficiency of the final test case. This synergy between generation and repair elevates the correctness and sufficiency of the produced test cases significantly beyond previous methods. 
Through comparative experiments, TestART demonstrates an 18\% improvement in pass rate and a 20\% enhancement in coverage across three types of datasets compared to baselines. 
Additionally, it achieves better coverage rates than EvoSuite with only half the number of test cases. 
These results demonstrate TestART's superior ability to produce high-quality unit test cases by harnessing the power of LLMs while overcoming their inherent flaws. 
\end{abstract}




\maketitle

\sloppy
\section{Introduction}
Unit testing is crucial in software development and is the basic procedure of the software testing process. The primary goal of unit testing is to generate a suite of unit test cases that can improve code coverage and detect errors in the targeted software early during the development process. In unit testing, the focal method, which generally refers to the specific method or function being tested, is the primary target of most unit tests.
However, manually creating and maintaining unit test cases can be laborious and time-consuming~\cite{AthenaTest}. To mitigate these difficulties, researchers propose various methodologies and tools to automate the unit test generation process. 
Based on different technological foundations and application scenarios, the automated unit test generation tools can be categorized into two main types~\cite{tang2024chatgpt}: traditional program analysis-based~\cite{fraser2011evosuite, lukasczyk2022pynguin} and language-model-based~\cite{A3test, AthenaTest, xie2023chatunitest, bhatia2023unit, yuan2023no}. Traditional program analysis-based methods include search-based~\cite{baresi2010testful, derakhshanfar2022basic, fraser2010mutation, harman2001search}, constraint-based~\cite{ma2015grt, sakti2014instance}, and random-based~\cite{andrews2011genetic, pacheco2007feedback} techniques. However, the most widely used search-based software testing (SBST) tools (e.g., EvoSuite~\cite{fraser2011evosuite} and Pynguin~\cite{lukasczyk2022pynguin}) generate test cases that differ significantly from human-written ones. This makes it challenging to read, understand, or modify to reuse the generated unit test cases~\cite{gargari2021sbst}. 
Language model-based approaches based on transformer architecture~\cite{vaswani2017attention} like A3Test~\cite{A3test} and AthenaTest~\cite{AthenaTest} can learn from real-world focal methods and generate developer-written test cases. However, they cannot even correct a syntax error by engaging with the model, resulting in most generated tests being incorrect and low passing rates. Recently, large language models (LLMs, e.g., ChatGPT~\cite{achiam2023gpt}) demonstrate improved performance over previous language-model-based methods on unit testing, represented by several works~\cite{xie2023chatunitest, bhatia2023unit, yuan2023no, alshahwan2024automated}. These methods can generate test cases that approximate human-level comprehension and deliver relatively high coverage, making them a central focus in this field.

However, these methods remain constrained by the instability and hallucination issues inherent in LLMs. While fine-tuning techniques have been proposed to mitigate hallucination problems, they tend to increase both time and financial costs. Additionally, ineffective prompt engineering, along with an insufficient understanding of environmental dependencies, can lead to compilation errors due to invalid context information. Furthermore, LLMs can barely run and analyze the generated test cases~\cite{xie2023chatunitest}, so they are hard to obtain the testing feedback, including error report or coverage feedback~\cite{tang2024chatgpt}, which may cause runtime errors and low coverage rates. Although developers can re-enter the feedback information from compilers to LLMs and request them to repair bugs, the inevitable faithfulness hallucination~\cite{huang2023survey} problem could make this self-repair interaction into potentially endless iterations. Even with continuous attempts, the test cases remain difficult to optimize due to the impact of the repetitive suppression problem~\cite{xu2022learning, zhang2023multi}.

To address the aforementioned issues and mitigate the impact of hallucinations, in this paper, we propose \textbf{TestART}.
TestART is a novel unit test generation method enhanced by the co-evolution of automated generation and repair iteration based on the LLMs (e.g., ChatGPT-3.5). We aim to utilize the generation capability of LLMs and the synergy of the generation-repair mechanism to create high-quality unit test cases. TestART starts by pre-processing the source code and then utilizes the LLMs to generate an initial set of test cases. These test cases undergo a loop involving compilation, execution and repair to identify, capture and fix errors based on TestART's repair strategy. TestART uses a template-driven repair strategy tailored to the outputs of LLMs, prioritizing template-based repairs and resorting to LLM-based repair only for unresolved cases. After that, TestART evaluates the coverage rates of executed test cases. If the coverage meets a high standard, the test cases are output as the final result. If not, we use the positive prompt injection and the coverage-guided testing feedback to alleviate the hallucination and incrementally generate test cases for the next iteration. As a result, TestART can automatically generate test cases with high passing execution and coverage rates under the co-evolution between the generation and repair processes, reducing manual intervention significantly.

To verify the effectiveness of TestART, we conduct extensive experiments to compare TestART with state-of-the-art automated unit test generation approaches on open-source and industrial benchmarks. The experimental results show that the pass rate of test cases generated by TestART can far exceed other methods, reaching 78.55\% on Defects4J, with an improvement of 18\% compared to both ChatGPT-4.0 model and ChatUniTest. In addition, TestART achieves an average line coverage rate of 90.96\% on passed focal methods, exceeding the SBST tool EvoSuite by 3.4\% with only half the number of test cases. We also complete ablation experiments to verify the effectiveness of modularization. The experimental results show that TestART effectively utilizes LLMs' generation capability and generates high-quality unit test cases.

To sum up, the main contributions of this paper are as follows:  
\begin{itemize}
\item \textbf{Method}. We propose TestART, a method designed to enhance LLM-based unit testing through the synchronized evolution of automated generation and iterative repair. TestART leverages the generative capabilities of LLMs by integrating generation-repair co-evolution, testing feedback and positive prompt injection into the iteration, repairing the bugs contained in the generated test cases and feeding back the coverage information for outputting high-quality test cases.  
\item \textbf{Tool}. We integrate TestART into a Python tool, which is the first tool that combines automated unit test generation with APR. We make the code for the tool and the experimental data available on GitHub (\url{https://github.com/sikygu/TestART}). 
\item \textbf{Study}. We conduct a comprehensive experiment to evaluate the performance of TestART on the open-source and industrial datasets. Compared with different baselines, TestART demonstrates superior performance, with an 18\% improvement in pass rate and a 20\% enhancement in coverage rate. Additionally, TestART achieves better coverage than EvoSuite while utilizing only half the number of test cases. The evaluation results demonstrate that TestART can suppress the hallucination problem and generate high-quality test cases.

\end{itemize}

\section{Background and Motivation}

\begin{figure*}[!tbp]
  \centering
  \includegraphics[width=0.9\linewidth]{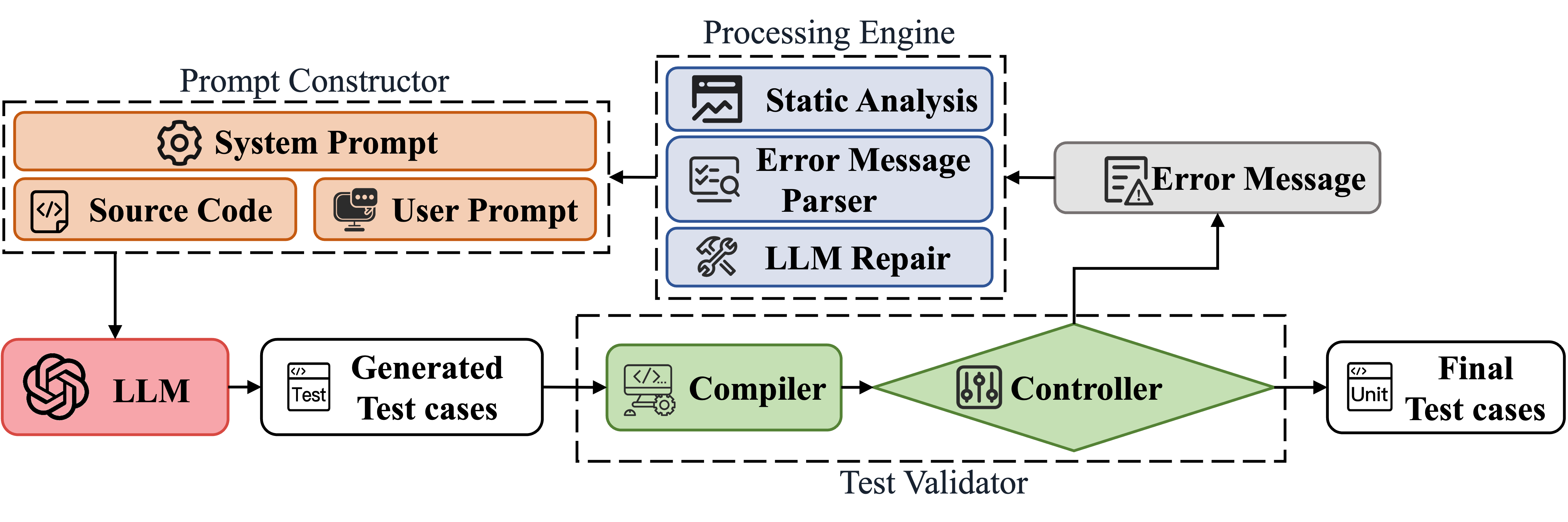}
  \caption{The common workflow of LLM-based unit testing generation method}
  \label{background}
\end{figure*}

\begin{figure}[!tbp]
  \centering
  \includegraphics[width=0.65\linewidth]{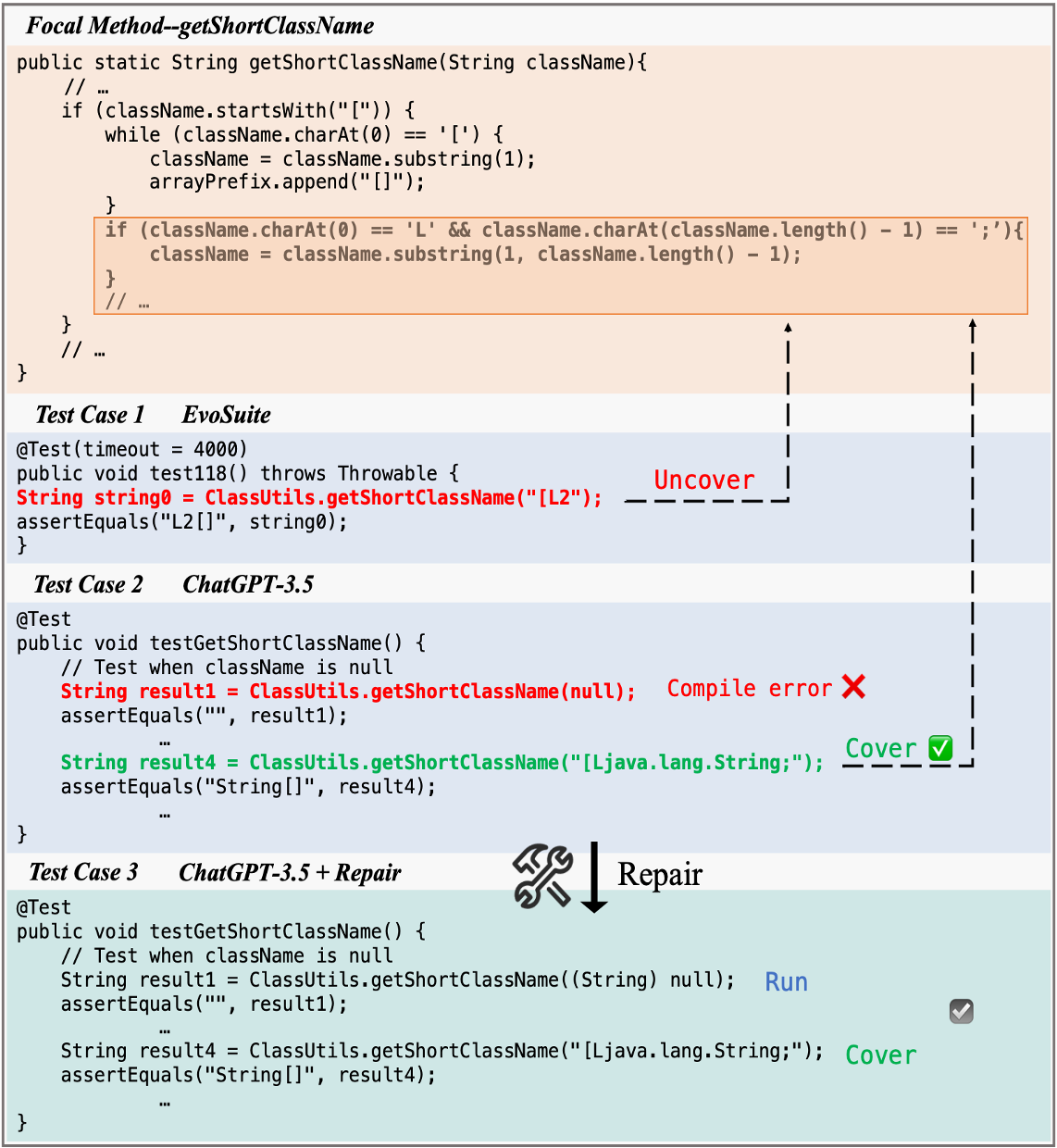}
  \caption{The motivation of TestART}
  \label{motivation}
\end{figure}

This section outlines the typical workflow for utilizing LLMs to generate unit test cases and presents an example to illustrate the motivation behind the proposal of TestART.

\subsection{LLM-based Unit Test Generation}
Nowadays, with the explosion of LLMs represented by ChatGPT, the LLM-based unit test generation methods attract more attention. To reduce time and financial costs, mainstream researchers reduced their reliance on fine-tuning strategies and instead shifted towards exploring viable prompting mechanisms for generating test cases~\cite{schafer2023adaptive, xie2023chatunitest,yuan2023no,wang2024hits}. Based on previous works, Fig.~\ref{background} illustrates the commonly used workflow for constructing prompts and performing testing validation to generate test cases using LLMs. Most of the approaches consist of three main modules: \textit{prompt constructor, test validator and processing engine}. Firstly, the \textit{prompt constructor} initializes system prompts, defining the role of the LLMs as a ``testing expert''. Then, the source code is taken as part of different user prompts, instructing LLMs to generate the test cases. Secondly, after getting the generated test cases, the \textit{test validator} invokes the compiler and controller to perform dynamic analysis on the test cases. If the test cases do not compile or execute successfully, they will be subject to the \textit{processing engine} along with the error message. Thirdly, researchers use various engines, including static analysis, error message parser, or LLM-repair, to further process the results and update the user prompt to improve the LLM-generated test cases. Ultimately, after several iterations, these methods yield test cases of comparable quality.

\subsection{Motivation}
To further elucidate the limitations of using LLMs, we present an example to introduce our motivation. As shown in Fig.~\ref{motivation}, a focal method \textit{getShortClassName} from project \textit{Lang} in dataset Defects4J is used to generate unit test cases. The function of the code snippet is to resolve the Java class name. EvoSuite and ChatGPT-3.5 are used to generate Test Case 1 and Test Case 2 for this focal method, respectively. After compiling and executing the two test cases, we inspect the testing results and the coverage area for the source code. Even though Test Case 1 can pass the compilation and execution, it cannot cover a deeper branch (the ``if'' branch framed in orange) due to the complexity of the preconditions. To satisfy the condition to enter this branch, the original \textit{className} string should be an internal representation indicating an array of reference data types (e.g., \textit{Ljava.lang.String}). Conversely, Test Case 2 seems to meet the condition for covering this branch but it fails to compile due to a simple error of variable type in another section of the code. In this instance, repairing the compilation error in Test Case 2 by changing the variable \textit{null} to \textit{(String) null} yields Test Case 3, which successfully runs and covers that branch.

The main insight from this example is that although LLM-generated test cases exhibit high logic quality, they are frequently compromised by simple errors, adversely affecting the pass and coverage rate. In fact, the pass rate of test cases directly generated by ChatGPT does not exceed 50\%~\cite{yuan2023no}. Meanwhile, due to the issue of repetitive suppression in LLMs~\cite{zhang2023multi}, the output from LLMs often includes repetitive or highly similar results, containing similar errors. Therefore, we propose applying the automated program repair (APR) technique that utilizes predefined templates to precisely and reliably correct these errors rather than only relying on LLMs to repair bugs.

\section{Approach} 
\begin{figure*}[!tbp]
  \centering
  \includegraphics[width=0.98\linewidth]{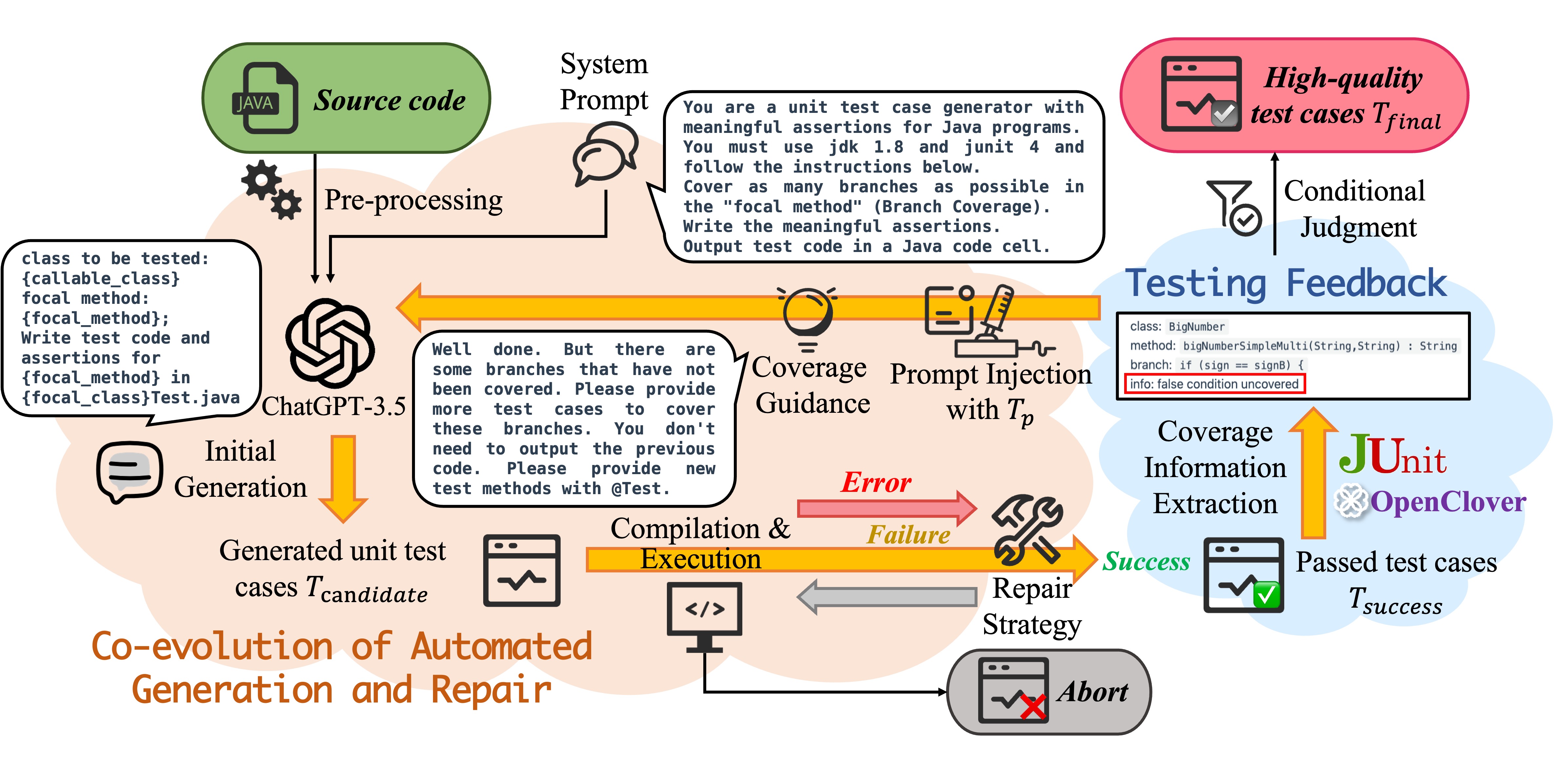}
  \caption{The Overview of TestART}
  \label{overview}
\end{figure*}

In this section, we first present the overview of the proposed TestART, followed by a detailed examination of its key modules. 
The workflow of TestART is illustrated in Fig.~\ref{overview}. 
Given a source code, TestART initially performs pre-processing (as described in Sec.~\ref{sec:approach_preprocess}) to alleviate the hallucination problem. 
Afterward, TestART leverages the ChatGPT-3.5 model to generate the initial set of test cases $T_{candidate}$ (as explained in Sec.~\ref{sec:appraoch_initial}). 
Once generated, $T_{candidate}$ enters the co-evolution loop (depicted by the yellow arrows in Fig.~\ref{overview}), which aims to produce high-quality final test cases. 
Specifically, TestART submits $T_{candidate}$ for compilation and execution to capture error messages. Based on the captured errors (compilation errors, runtime errors, or other detected bugs), TestART applies the repair strategy to address these errors and recompiles and runs until it passes without bugs or reaches the maximum number of iterations. 
If the test cases still unsuccess, $T_{candidate}$ is discarded. Otherwise, $T_{candidate}$ is promoted to $T_{success}$. Next, TestART employs JUnit and OpenClover to evaluate the coverage of $T_{success}$ over the source code, transforming the uncovered areas into test feedback. 
If $T_{success}$ meets the coverage standard, it will be output as the final result ($T_{final}$). Otherwise, $T_{success}$ and the coverage-guidance feedback are provided as positive prompt injections to the ChatGPT-3.5 model, continuing to the next co-evolution loop and revert to $T_{candidate}$. 
A detailed description of the co-evolution of automated generation and repair is introduced in Sec.~\ref{sec:appraoch_synergy}. 
Benefiting from this synergy, TestART can iterate test cases incrementally and ensure that each round of test cases passes successfully, which continuously improves coverage.

\subsection{Pre-processing}
\label{sec:approach_preprocess}

To alleviate the LLM hallucination issue and reduce token consumption, the source code undergoes a pre-processing step, including removal and compression. TestART first removes comments and redundant blank lines from the code. Comments may make the code bulky, which is not conducive to maintenance and debugging and is prone to the LLM hallucination problem caused by inconsistencies between comments and code content. Eliminating extra blank lines helps standardize the code. Afterward, TestART compresses the excessive source code context. Secondly, LLMs do not perform well in processing overly long texts (although they do not exceed the token limit), but the focal method and its callable functions may be extensive. This phenomenon brings two main issues: 1) the callable classes and functions have lengthy text in total, and simple deletion can affect the accuracy of the test case; 2) a large amount of invalid content leads to an excess of redundant information. Therefore, designing a compression strategy that maintains the code at an optimal length is essential, enabling LLMs to focus on the focal method. Considering that the method body accounts for a large proportion of the entire class structure, TestART's compression strategy focuses on compressing callable functions by only including method signatures while retaining the complete method bodies of focal methods. Following is an example:
\\

\begin{lstlisting}[linewidth=0.98\textwidth, language=Java]
-   public static float toFloat(final String str){
        return toFloat(str, 0.0f);
-   }
+   public static toFloat(String): float
\end{lstlisting}

It is worth noting that other elements, such as class variables and constants, are not compressed in this process to preserve the key structure and functionality of the code. Thus, the compressed context can still clearly describe the test scenario while maintaining a minimal token. In all, our compression strategy aims to reduce irrelevant information interference while retaining enough context to support the next generation step.

\subsection{Initial Generation}
\label{sec:appraoch_initial}

After pre-processing the source code, TestART utilizes the system prompt and fills the initial generation prompt templates to generate the initial candidate test cases. The system initialization prompt is depicted in the system prompt part at the top of Fig.~\ref{overview}. To make LLMs understand the testing environment and reduce the token consumption, TestART extracts the variables required for running in focal methods, including method signatures, the number of lines starting and ending the method, etc. After that, TestART produces the test code and related information to fill the initial generation prompt templates (far left of Fig.~\ref{overview}). TestART requests the LLMs to process the filled prompt and generate the initial $T_{candidate}$. This enables TestART to transform the initial set of testing requirements into tasks that LLMs can easily comprehend, thereby reducing human-computer interaction and achieving automated test case generation. Then, TestART tries to compile and run the generated test cases.

\subsection{Synergy}
\label{sec:appraoch_synergy}

However, the initial $T_{candidate}$ are imperfect, leading to low pass and coverage rates. Therefore, we propose the co-evolution of automated generation and repair iteration to iterate the test cases. The synergy process is reflected in 1) the repair strategy, which mainly uses templates to precisely repair the errors in $T_{candidate}$ to get $T_{success}$, and 2) the generation process, which uses positive prompt injection and coverage guidance to incrementally generate next iteration $T_{candidate}$ based on $T_{success}$. This section details the Repair Strategy, Coverage-guided Testing Feedback, and Positive Prompt Injection in the following.

\subsubsection{Repair Strategy}

By analyzing the errors in the initial $T_{candidate}$, we find that the LLM-generated test cases exhibit systematic and distinct errors. Moreover, due to the problem of repetitive suppression, these errors tend to recur. Because these errors are inherently produced by the LLMs, correcting them using the models alone proves challenging. Consequently, we developed repair templates based on expert knowledge, drawing upon the most advanced traditional repair techniques. Specifically, TestART first identifies the location of the erroneous code and extracts the error messages. We separate the compilation and execution of the test code into two distinct steps to more accurately identify the source of any errors and apply the corresponding repair templates. For compilation errors, TestART identifies issues by analyzing the logs generated during compilation. These logs contain detailed error messages, the location in the code where the error occurred, and relevant warning information. For runtime errors, TestART locates errors by examining the stack trace information. Then TestART utilizes corresponding templates to repair the test code. To address spelling and range errors that template-based repair cannot cover, TestART attempts a one-time LLM-based repair after multiple failed template repair attempts. Unlike previous repair strategies applied to semantic buggy code~\cite{liu2019tbar,zhang2023gamma,liu2019avatar}, TestART focuses on repairing compilation errors (including syntax errors, import errors, and scope errors) and runtime errors that appear in the generated test cases. Our repair strategy focuses more on the internal logic and expected behavior of the test cases, aiming at increasing the correctness. TestART can repair the most common errors in LLM-generated test cases by designing five experiential repair templates, which are simple, stable and effective (reaching 50\% and 75\% of compile and runtime error repair rates, respectively). In the following, we demonstrate the five templates and how the buggy test code can be repaired using the designed templates.

\textbf{T1: Check Package Import.} The most common compilation error is the symbol parsing error, which occurs when the compiler generates a ``cannot find symbol'' prompt. Import errors often account for a large proportion of the time. TestART first indexes the test project, JDK and all third-party dependent JAR packages to obtain the fully qualified class names of all accessible Java classes (e.g., java.util.HashMap) during testing. During compilation, determine whether an import error causes the compilation error. If so, extract the missing class name from the compilation result and find the fully qualified class name of the class from the index, then import it. In T1, \textit{ClassName} is an unimported class, and \textit{packageReference} is the package name where the class is located. 

\begin{lstlisting}[linewidth=0.98\textwidth, language=Java]
+   import packageReference.ClassName;
\end{lstlisting}

After compilation, test cases fail when the test results do not match expectations, which is called test failure. We design templates for two main situations: assertion error and runtime error. If ``AssertionError'' or ``org.junit.ComparisonFailure'' is found in the stack trace information, TestART checks the assertion type and error code line, using the corresponding template to repair it.

\textbf{T2: Mutate Assertion Statements}. Boolean assertion errors comprise a significant proportion of runtime errors (accounting for 25\% in our experiments). When the \textit{assertNull} method is used in a test case for assertion and the test fails, it means that the object being checked is not null, which contradicts the expectation. In such cases, correcting \textit{assertNull} to \textit{assertNotNull} is a quick-fix strategy. Conversely, if \textit{assertNotNull} is used and the test fails, it indicates that the object is null. In this case, \textit{assertNotNull} should be changed to \textit{assertNull}. This template is also applied to \textit{assertTrue} and \textit{assertFalse}. \\

\begin{lstlisting}[linewidth=0.98\textwidth, language=Java]
-   Assert.assertNull(param);
+   Assert.assertNotNull(param);
or
-   Assert.assertNotNull(param);
+   Assert.assertNull(param);
\end{lstlisting}

\begin{lstlisting}[linewidth=0.98\textwidth, language=Java]
-   Assert.assertTrue(param);
+   Assert.assertFalse(param);
or
-   Assert.assertFalse(param);
+   Assert.assertTrue(param);
\end{lstlisting}

\textbf{T3: Replace Expected Values}. AssertEqual assertion errors account for the highest proportion of runtime errors (accounting for 40\% in our experiments). When using the \textit{assertEquals} method in test cases for assertion, it is usually necessary to compare whether two values are equal. If the test fails, the reason is the expected value does not match the actual execution result. In some cases, the expected value in the test case is incorrectly specified. T3 replaces the expected value in \textit{assertEquals} from an incorrect or outdated value to the correct current actual value. We use regular expressions to extract the expected and actual values from the test report. To maintain the intended meaning of assertions and verify the expected output of expressions, T3 does not directly switch from using \textit{assertEquals} to \textit{assertNotEquals}. 

\begin{lstlisting}[linewidth=0.98\textwidth, language=Java]
-   Assert.assertEquals(expectedValue, expression);
+   Assert.assertEquals(actualValue, expression);
\end{lstlisting}

\textbf{T4: Insert Check Statements}. Unit test cases often define the test oracle through exceptions. When running unit tests, if a test case encounters a runtime exception, it usually means there is a potential error in the code or the test case fails to simulate exception handling logic correctly. T4 is based on the type of error ExceptionType thrown by the target code line, wrapping the error code line with try-catch statements and catching the corresponding exception.

\begin{lstlisting}[linewidth=0.98\textwidth, language=Java]
-   obj.method1();
+   try{
        obj.method1();
    }catch(ExceptionType e){
    // Expected
+   }
\end{lstlisting}

\textbf{T5: Mutate Check Statements}. In the LLM-generated unit test cases, a certain code segment may throw an exception and thus add a try-catch statement to catch the expected exception type. However, suppose the caught exception type is incorrect. In that case, the actual thrown exception type mismatches the specified exception type in the catch statement, and the buggy code causes an uncaught exception at runtime, potentially leading to program crashes or unstable operations. In addition, a code segment can throw different types of exceptions. Merely catching one type of exception is insufficient to thoroughly handle all potential error situations. Therefore, if the existing catch statement does not catch the actual thrown exception type, T5 adds a new catch statement specifically for this newly discovered exception type. This template ensures that the handling logic for the original exception types remains intact to avoid introducing new errors from modifications and covers a broader range of error scenarios.

\begin{lstlisting}[linewidth=0.98\textwidth, language=Java]
    try{
        obj.method1();
    }catch(ExceptionType1 e){
        //mismatched or insufficient 
    }
+   catch(ExceptionType2 e){
        // Expected
+   }
\end{lstlisting}

\begin{figure*}[!htb]
  \centering
  \includegraphics[width=0.95\linewidth]{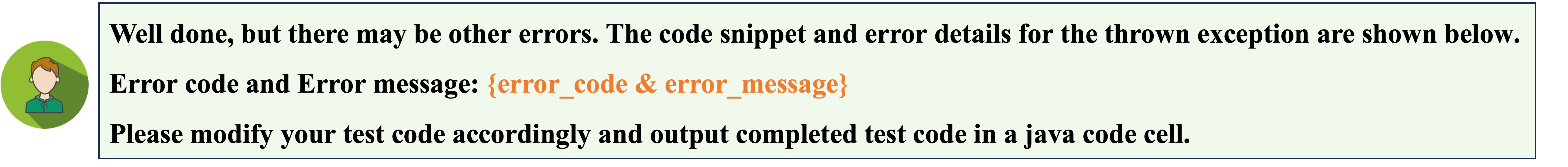}
  \caption{The prompt of LLM-based repair}
  \label{repair_prompt}
\end{figure*}

\textbf{LLM-based Repair.} The template repair cannot fix the remaining detail errors, such as syntax and range errors. After trying to match and repair using the five templates, if there is still a situation that fails to compile or run, TestART uses the corresponding prompt to guide the LLMs to repair the test code; the prompt information is as shown in Fig.~\ref{repair_prompt}. Finally, successfully repaired $T_{candidate}$ becomes $T_{success}$ and feeds to the next module.

\subsubsection{Coverage-guided Testing Feedback}

Since LLMs can barely validate $T_{success}$, testing sufficiency is hard to guarantee and improve. Therefore, TestART proposes the coverage-guided testing feedback module to guide LLMs in generating test cases based on uncovered branch conditions iteratively. Specifically, after getting $T_{success}$ in the current round, according to Fig.~\ref{overview}, TestART invokes the Junit and OpenClover tools to calculate the coverage of the executed test cases and provide coverage-guided testing feedback.
First, if the branch coverage reaches the coverage standard (e.g., 95\%) or the number of iterations reaches the preset limit, TestART outputs the current test case as a high-quality test case and ends the iteration. If not, second, TestART calculates the code coverage by identifying branches in the focal method not covered by testing. Afterward, TestART processes this information and provides feedback to LLMs to guide the generation in subsequent rounds. Specifically, TestART extracts all uncovered branch content and integrates the testing information into a coverage report. This report includes class names, method names, and the specific code of the uncovered branches, along with an indication of whether the true or false branches are not covered. Third, the testing information is formatted according to a predefined template (shown in Fig.~\ref{prompt}) and sent to the LLMs.

\subsubsection{Positive Prompt Injection}

\begin{figure}[!tb]
  \centering
  \includegraphics[width=0.65\linewidth]{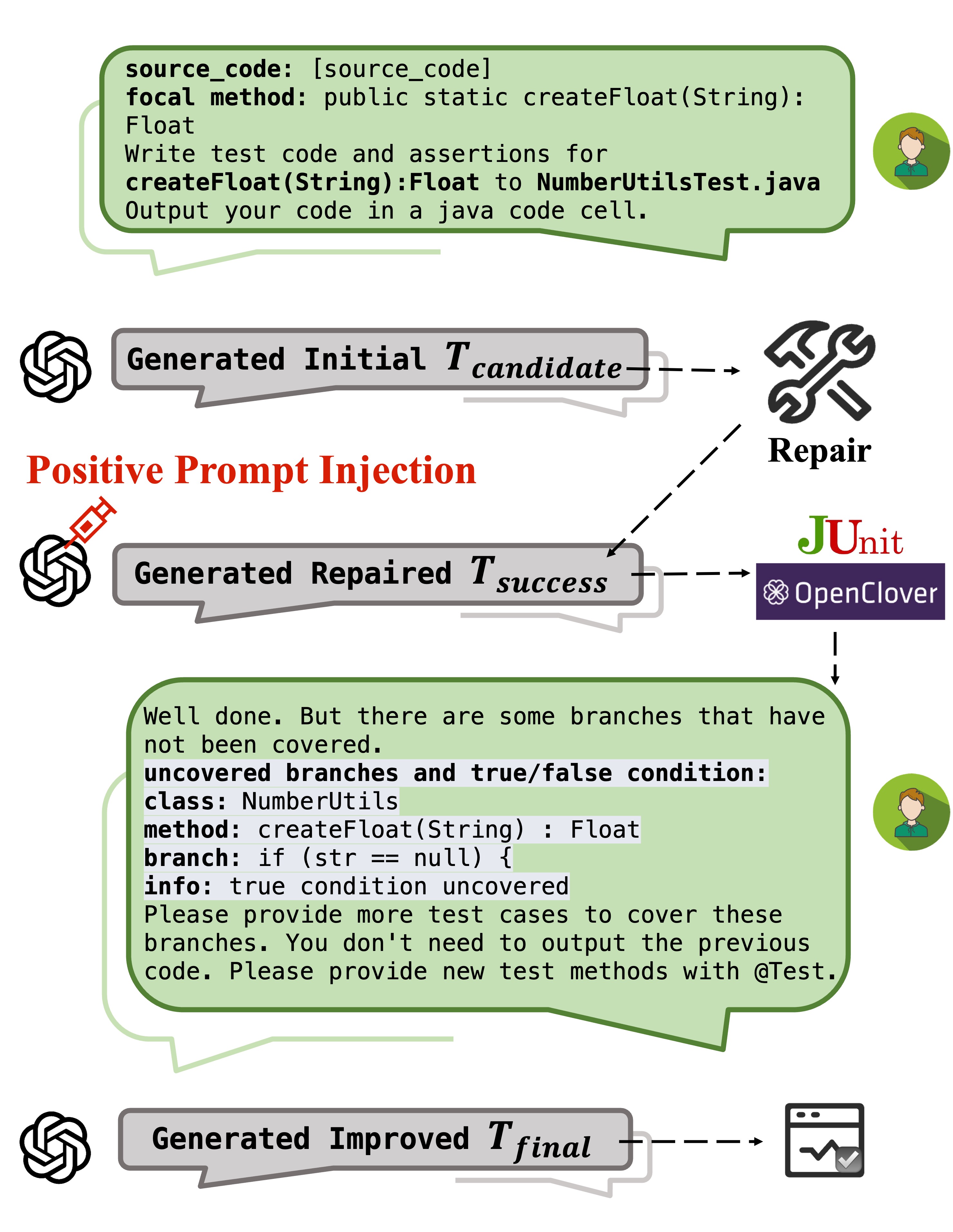}
  \caption{The process of positive prompt injection and coverage-guided testing feedback}
  \label{prompt}
\end{figure}

While TestART repairs the initial $T_{candidate}$ to obtain $T_{success}$ and extracts feedback on test coverage, LLMs remain unaware of these repair actions when $T_{success}$ enters the next iteration of co-evolution to increase coverage. Consequently, in subsequent iterations, the LLMs continue generating based on the initial $T_{candidate}$ instead of the improved $T_{success}$. Due to the hallucination problem, directly informing the LLMs that $T_{success}$ is the revised version still results in errors in subsequent outputs. To alleviate this situation, we design a positive prompt injection module to help LLMs generate better incremental test cases in the co-evolution process and reduce the required tokens.

Prompt injection technique is used for attacking and defending LLMs~\cite{choi2022prompt}. It involves inserting unexpected or malicious content into the prompt to change the system's behavior, extract unauthorized data, or trigger unintended actions. The core is that altering the output of the LLMs induces LLMs to perceive the modified content as its original generation. However, unlike injecting attacks, which are considered negative, TestART treats the repaired test cases $T_{success}$ as positive injections to mitigate hallucination.

Precisely, TestART injects ($T_{success}$) as a positive prompt to minimize context and computational costs. This module enables LLMs to consistently generate the same or similar test cases within the correct paradigm. As shown in Fig.~\ref{prompt}, it indicates one iterative process of generation, repair, positive prompt injection and coverage-guided testing feedback. After initial generation, LLMs output the generated unit test cases $T_{candidate}$, which may include several bugs. Then TestART applies the repair strategy and successfully repairs the bugs, resulting in $T_{success}$. To mitigate hallucination issues, TestART injects $T_{success}$ as the prompt in place of $T_{candidate}$, leading LLMs to perceive $T_{success}$ as the initial generation. Next, TestART calculates the coverage information and constructs the template to give feedback to LLMs. Through incremental evolutionary iterations, $T_{success}$ achieves the coverage standard and outputs as $T_{final}$.

\section{Experiment Design}

Our evaluation is designed to answer the four main research questions (RQs):

\textbf{RQ1:} How does the correctness of the test case of TestART compare to the baseline? What are the error types in the test cases and how well does TestART repair different error types?

\textbf{RQ2:} How does the sufficiency of the test case of TestART compare to the baseline? 

\textbf{RQ3:} How does the combination of different parts impact the robustness of TestART?

\textbf{RQ4:} What is the performance of TestART when applied to unlearned datasets? and how well does it generalize to new data?

\textbf{RQ5}: What is the time and money cost of TestART to generate test cases?

\subsection{Datasets}

\begin{table*}
\centering
\caption{The description of datasets used in the experiment}
\label{dataset}
\resizebox{\linewidth}{!}{
\begin{tabular}{ccccccc} 
\toprule
Dataset                    & Project name             & Abbr. & Version & Focal methods & Total & Learned?  \\ 
\midrule
\multirow{5}{*}{Defects4J} & Gson                     & Gson  & 2.10.1  & 378           &  \multirow{5}{*}{8192} & Y    \\
                           & Commons-Lang             & Lang  & 3.1.0   & 1728     &     & Y         \\
                           & Commons-Cli              & Cli   & 1.6.0   & 177       &    & Y         \\
                           & Commons-Csv              & Csv   & 1.10.0  & 137       &    & Y         \\
                           & JFreeChart               & Chart & 1.5.4   & 5772       &   & Y         \\ \midrule
\multirow{3}{*}{HITS dataset}      & Event-ruler              & RUL   & 1.4.0   &  165       &  \multirow{3}{*}{885}      & N         \\
                           & Windward                 & WIN   & 1.5.1   &  156        &    & N         \\
                           & batch-processing-gateway & BPG   & 1.1     &  564         &   & N         \\ \midrule
\multirow{4}{*}{Internal dataset}   & Project1 & P1   & -       & 82     &  \multirow{4}{*}{668}       & N   \\
                           & Project2 & P2    & -       & 332         &  & N         \\
                           & Project3 & P3   & -       & 36          &  & N         \\
                           & Project4 & P4   & -       & 218        &   & N         \\ 
\bottomrule
\end{tabular}
}
\end{table*}

To verify our method, we select the most commonly used open-source dataset Defects4J. We also use an unlearned dataset (created after the cutoff date of the training data of GPT-turbo-3.5) and an industrial dataset to avoid potential problems caused by data leakage. We extract all public, non-abstract classes as \textit{focal classes} (i.e., classes whose modifiers include ``public'' and do not include ``abstract''), and we extract public methods within these classes as \textit{focal methods} (i.e., methods that include the ``public'' modifier).

\textit{Defects4J}~\cite{just2014defects4j} is a collection of reproducible software bugs and supporting infrastructure to advance research in software engineering. We follow previous work~\cite{xie2023chatunitest} and select five Java projects from Defects4J to evaluate TestART, shown in Table~\ref{dataset}. We extracted public and non-abstract classes from five projects as focal classes and extracted their public methods as focal methods, a total of 8192 focal methods extracted.

\textit{HITS dataset}~\cite{wang2024hits} comprises internet-sourced projects developed after the knowledge cutoff date for ChatGPT-3.5, indicating that the LLMs have no training data from this dataset. The HITS dataset contains 885 focal methods from 3 projects. We retain the part of the dataset in the original paper that is not duplicated in Defects4J as this dataset.

\textit{Internal dataset} is an internal Java unit test dataset provided by Huawei. This dataset contains a total of 668 completely internal Java focal methods. This means that the open-source model cannot learn the data features of this part. The specific configuration of the dataset is shown in the Table~\ref{dataset}.

\subsection{Baselines}
To evaluate the effectiveness of our proposal, we compare TestART with four baselines. We choose the state-of-the-art methods of four kinds of solutions as baselines including the SBST tool (EvoSuite), the language-model-based method (A3Test), the Large language model (two ChatGPT models) and the LLM-based test case generation approach (ChatUniTest). Note that CHATTESTER is excluded from baseline comparisons due to 1) the similar LLM-based repair methodology with ChatUnitTest and 2) ChatUnitTest shows better performance than CHATTESTER in the experiments according to HITS\cite{wang2024hits}.

\textit{EvoSuite}~\cite{fraser2011evosuite} is a traditional unit test generation tool that uses evolutionary algorithms to create new test cases and a fitness function to direct the search process. This approach helps achieve high code coverage criteria, such as branch and line coverage.

\textit{A3Test}~\cite{A3test} is a test case generation approach based on deep learning that is enhanced with assertion knowledge and includes a mechanism to verify naming consistency and test signatures. A3Test applies domain adaptation principles to adapt existing knowledge from an assertion generation task to the test case generation task.

\textit{ChatGPT}~\cite{achiam2023gpt} is a highly advanced technology that can replicate human speech and reasoning by learning from a vast library of human communication. It achieves performance levels comparable to humans in professional and academic settings. ChatGPT-3.5 and ChatGPT-4.0 are set as two different baselines.

\textit{ChatUniTest}~\cite{xie2023chatunitest} is a ChatGPT-based automated unit test generation tool developed by the Generation-Validation-Repair framework. ChatUniTest generates tests by analyzing the project, extracting critical information, and creating an adaptive context that includes the focal method and its dependencies within a set token limit.

\subsection{Evaluation Metrics}
To validate the experimental results based on four research questions, we set four key evaluation perspectives: correctness (1-4), sufficiency (5-8), error detection (9-10), and the test case count (11). The specific descriptions are as follows:
\begin{enumerate}[leftmargin=2em]
\item \textbf{Syntax Error (SE)} refers to the percentage of the test code that includes Java syntax error, verifying by Java parser.
\item \textbf{Compile Error (CE)} refers to the percentage of the test code that produces errors during the compilation. 
\item \textbf{Runtime Error (RE)} refers to the percentage of the test code that includes error or failure during the execution.
\item \textbf{Pass Rate (Pass)} refers to the percentage of the test code that is syntactically accurate, compiles and runs without errors or failures. It includes calls to the tested method and assertions and allows the target focal method to pass.
\item \textbf{Branch Coverage of Correct Test (BCCT)} represents the branch coverage rate of the passed focal methods.
\item \textbf{Line Coverage of Correct Test (LCCT)} represents the line coverage rate of the passed focal methods.
\item \textbf{Total Branch Coverage (TBC)} represents the branch coverage ratio of all the focal methods.
\item \textbf{Total Line Coverage (TLC)} represents the line coverage ratio of all the focal methods.
\item \textbf{Mutation Coverage (MC)} represents the ratio of the number of mutations killed by tests to all the number of mutations (regardless of whether tests covered it or not).
\item \textbf{Test Strength (TS)} represents the ratio of the number of mutations killed by tests to the number of all mutations covered by tests.
\item \textbf{Test Case Count (TCC)} represents the total number of test cases generated by the testing approach.
\item \textbf{Assertion Count (AC)} represents the total number of assertions contained in the generated test cases.
\end{enumerate}

\subsection{Experimental Setup}

In the experiment, TestART generates unit tests for each focal method through up to \textbf{four iterations}.TestART selects the best test case by prioritizing execution success, maximum coverage, and minimal test count. To compare more fairly with the baseline, TestART and ChatUniTest utilize GPT-3.5-turbo-0125 as the basic model by calling the API, which offers a 16k context length, with the \textit{temperature} setting adjusted to 0.5. During the operation of ChatUniTest, we set its \textit{maxPromptTokens} parameter to 16,385 and generate unit test cases through the default \textit{maxRounds} of five iterations per attempt. When we conduct the baseline experiment using ChatGPT-3.5 and ChatGPT-4.0, we obtain test cases using the initial generation of TestART. We train the model of A3Test based on the Methods2Test dataset~\cite{tufano2020unit} and pre-trained model~\cite{ahmad2021unified}, setting the learning rate of 1e-5 for 110 epochs. We configure EvoSuite with 3 CPU cores, allocating 2000MB of memory to each core, and set the search time up to 10 minutes. During the testing process, Java 1.8 is used as the compiler and runtime environment, JUnit 4 is employed as the unit testing framework and OpenClover is used to calculate coverage rates. We calculate the mutation coverage based on the state-of-the-art mutation testing system PITest\footnote{https://github.com/hcoles/pitest.git}, and the mutators are set by default (11 kinds of mutators). We repeat each experiment three times and take the average as the final result.

\section{Results and Analysis}
In this section, we aim to evaluate the proposed TestART's performance based on the answers to all research questions.

\subsection{Answer to RQ1 (Correctness)} 

\begin{table*}
\centering
\caption{The correctness performance of TestART compared with different baselines on Defects4J}
\label{rq1}
\begin{tabular}{cccccccc} 
\toprule
Method  & Projects & Focal methods & Fail   & SE$\downarrow$ & CE$\downarrow$ & RE$\downarrow$ & Pass$\uparrow$  \\ 
\midrule
\multirow{5}*{TestART(Ours)}
 & Gson & 378 & 1.06\% & 0.79\% & 23.54\% & 11.11\% & 63.49\%  \\
& Lang & 1728 & 0.00\% & 0.12\% & 4.17\%  & 7.87\%  & 87.85\%  \\
& Cli  & 177  & 0.00\% & 1.69\% & 12.99\% & 9.60\%  & 75.71\%  \\
& Csv & 137   & 0.73\% & 0.00\% & 10.22\% & 12.41\% & 76.64\%  \\
& Chart & 5772 & 0.02\% & 0.50\% & 14.21\% & 8.39\% & 76.89\%  \\ 
\midrule
A3Test  &   &  & 0.00\%  & 24.43\%   & 34.77\%   & 25.54\%   & 15.26\%  \\ 

ChatGPT-3.5 &   &  & 0.07\% & 0.45\% & 25.09\% & 24.48\% & 49.91\%  \\ 

ChatGPT-4.0 &Total &8192 & 0.01\% & \textbf{0.05\%} & 19.43\% & 20.75\% & 59.75\% \\ 

ChatUnitTest & &  &0.70\%  &0.60\% & 23.35\%  & 16.47\% & 60.05\% \\

TestART(Ours) & & & 0.07\% & 0.45\% & \textbf{12.43\%} & \textbf{8.50\%} & \textbf{78.55\%} \\
\bottomrule
\end{tabular}
\end{table*}

\begin{figure*}[!tbp]
  \centering
 \includegraphics[width=\linewidth]{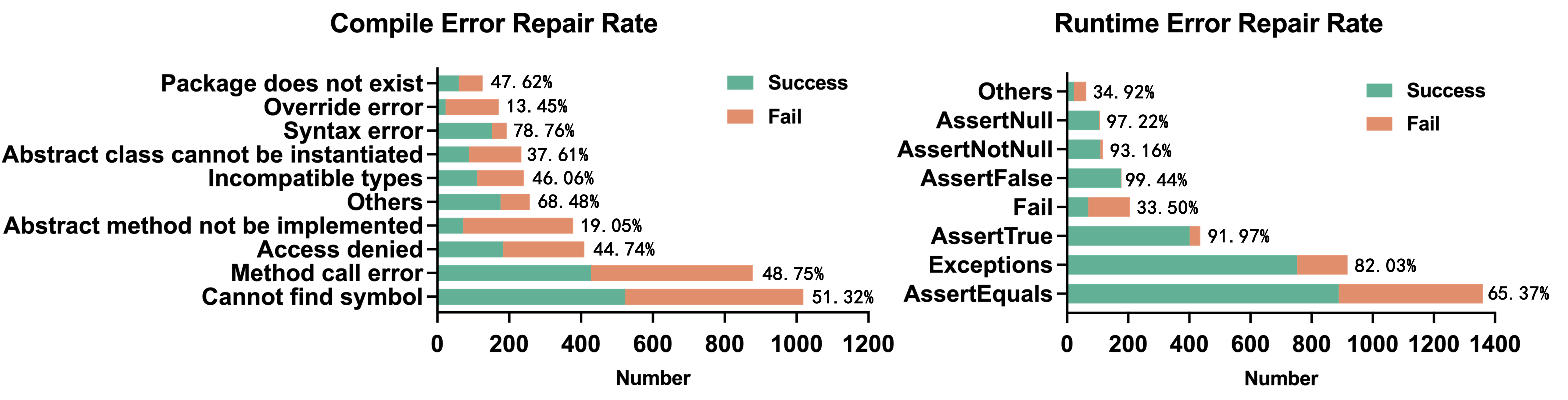}
  \caption{Repair rates of TestART on Defects4J}
  \label{RQ1}
\end{figure*}

\textbf{Design.}
We set RQ1 to determine the fundamental attribute of unit test cases: whether they pass when executed. The focal method is seen as the unit of calculation for all statistical results. If an approach fails to generate a unit test case for a specific focal method, that focal method is classified as ``Fail''. Otherwise, four metrics of correctness are calculated for the test cases of that focal method. In addition, we statistic the repair rate of all compile and runtime errors on different error types. EvoSuite primarily uses test cases as the unit of analysis for statistics, which differs from the setting of this experiment; therefore, we do not make comparisons with EvoSuite on correctness.

\textbf{Results.} 
As presented in Table~\ref{rq1}, we first demonstrate the results of our TestART on five sub-projects respectively. Then, we compare the average results of TestART and four baseline methods on the complete dataset.\textbf{ The data in the table clearly shows that TestART achieves the best results with a compilation error rate of 12.43\%, a runtime error rate of 8.50\%, and a final pass rate of 78.55\%}. Compared with the second place in each indicator, TestART reduces the compilation error rate by 7.0\% compared to ChatGPT-4.0 and was 8.0\% lower than ChatUniTest in the running error rate. The total pass rate is 63.29\% higher than A3Test, 28.64\% higher than ChatGPT-3.5, 18.80\% higher than ChatGPT-4.0, and 18.50\% higher than ChatUniTest. We also show the repair rates and the corresponding error types in Fig.~\ref{RQ1}. The three most common compilation errors (``cannot find symbol'', ``Method call error'' and ``Access denied'') can be repaired with about 50\% success. The overall repair rate of runtime errors is as high as 75\%.

\textbf{Analysis.} 
The experimental results show that TestART achieves the highest correctness due to the repair strategy. Although ChatUniTest also incorporates repair steps (primarily based on LLMs), experimental results demonstrate that using fixed repair templates in TestART is more effective. The insight is that the test cases generated by LLMs often have relatively consistent errors because of the repetitive suppression problem. However, LLMs can barely run the test cases to get error feedback, so using LLMs for debugging and repairing often gets stuck in a vortex, making it hard to achieve the passing execution results. Further, to specifically demonstrate the error types fixed by TestART and how well it repairs, we present the repair rates of ten compile error types and eight runtime error types in Fig.~\ref{RQ1}. Error types that occur less than 50 times are classified as ``other''. The implementation and inheritance of the abstract method and class lead to the lowest compile error repair rate for ``Abstract method not be implemented'' and ``Abstract class cannot be instantiated'' errors. The two errors with the highest compile repair rate are ``Syntax error'' and ``Cannot find symbol'', which benefit from feedback error reports to LLMs for self-repair and \textbf{T1} in TestART, respectively. Relatively, the runtime error repair rate is much higher than the compile error. \textbf{T2} in TestART makes four error types (``AssertNull'', ``AssertNotNull'', ``AssertTrue'' and ``AssertNotNull'') almost 100\% repaired. Due to complex parameter formats and data types, errors caused by ``AssertEquals'' and ``Exceptions'' can still be repaired by 65.37\% and 82.03\% based on \textbf{T3} and \textbf{T4}, respectively. Note that the errors of ``Fail'' and ``Others'' are repaired by LLMs, so the repair rates are relatively low. 

\begin{tcolorbox}[colback=gray!10, colframe=black, title=Answer to RQ1, boxrule=0.5mm, left=0mm, right=0mm, top=0mm, bottom=0mm]
\textbf{TestART outperforms all the baselines on the evaluation of correctness. The repair strategy effectively repairs main compile and runtime errors, improving the quality and usability of LLM-generated test cases.}
\end{tcolorbox}

\begin{table}[!tbp]
    \centering
    \caption{Total coverage comparison on Defects4J}
    \label{rq2}
    \begin{tabular}{ccc} 
      \toprule
      Method  &  TBC$\uparrow$ & TLC$\uparrow$   \\ 
      \midrule
      A3Test & 15.04\%   &  14.63\%         \\
      ChatGPT-3.5  &   43.10\%   & 42.58\%   \\
      ChatUnitTest &    48.68\%  & 47.39\%   \\
      ChatGPT-4.0  &   51.86\%   & 50.88\%   \\
      TestART(Ours) &   \textbf{69.40\%}  & \textbf{68.17\%}   \\
      \bottomrule
    \end{tabular}
\end{table}
\begin{table}[!tbp]
    \centering
    \caption{The ablation study results of TestART on Defects4J}
    \label{rq3}
    \begin{tabular}{cccc} 
      \toprule
      Method                & TBC $\uparrow$ & TLC $\uparrow$ & Pass$\uparrow$     \\
      \midrule
      ChatGPT-3.5 & 43.10\%  & 42.58\%  & 49.91\%  \\
      + Repair     & 62.13\%   & 62.24\%   & \textbf{78.55\%}  \\
      + Repair + Iteration  & 66.48\%   & 64.49\% & \textbf{78.55\%}  \\
      TestART & \textbf{69.40\%} & \textbf{68.17\%} & \textbf{78.55\%}  \\
      \bottomrule
    \end{tabular}
\end{table}

\begin{figure*}[!tbp]
  \centering
  \includegraphics[width=0.95\linewidth]{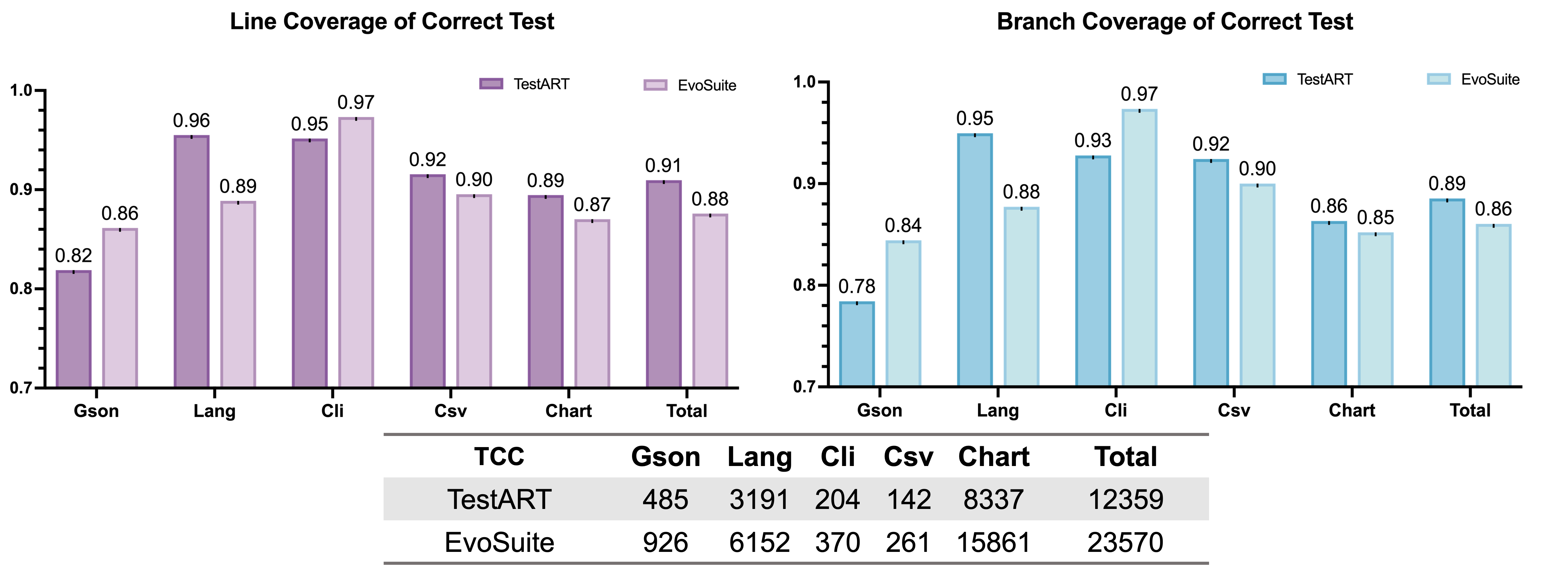}
  \caption{Coverage rate of correct test and test case count for TestART and EvoSuite on Defects4J dataset.}
  \label{RQ2-F}
\end{figure*}

\subsection{Answer to RQ2 (Sufficiency)}
\textbf{Design.}
To answer this research question, we compare the coverage of unit tests generated by TestART and the other four baselines. The coverage of the source code presents the sufficiency of testing. We use four main evaluation metrics to describe the sufficiency of the generated unit test cases. We count the number of focal methods for which passing test cases are generated by different methods (passed focal methods). Due to this difference, we calculate two coverage metrics (branch and line) based on the passed focal methods and the total focal methods, respectively. Total coverage indicates the testing results for the overall dataset, while the coverage of correct tests reflects the quality of coverage of test cases, both contributing to testing sufficiency.

\textbf{Results.} 
Table~\ref{rq2} and Fig.~\ref{RQ2-F} present the sufficiency of TestART compared to the baselines. As shown in Table~\ref{rq2}, \textbf{TestART achieves the highest TBC and TLC, which are 69.40\% and 68.17\%, respectively.} Compared to the second-ranked ChatGPT-4.0, TestART improves TBC by 17.54\% and TLC by 17.29\%. As the same method experimented with based on the ChatGPT-3.5, ChatUniTest is not superior to ChatGPT-4.0 but exceeds the coverage of ChatGPT-3.5. Meanwhile, TestART achieves average coverage rates of 17\% higher than ChatGPT-4.0. Fig.~\ref{RQ2-F} shows BCCT, LCCT and the generated test case count of TestART and EvoSuite for five sub-projects. \textbf{TestART achieves 88.54\% branch coverage and 90.96\% line coverage rates, exceeding EvoSuite by 2.5\% and 3.4\%, respectively.} It is noteworthy that the test cases generated by TestART are about \textbf{half} the number produced by EvoSuite (12,359 compared to 23,570), greatly reducing the execution costs.

\textbf{Analysis.}
Through Table~\ref{rq2}, we can see that TestART achieves the unquestionably highest coverage results compared to LLM-based methods on the total dataset. This improvement mainly benefited from the coverage-guided testing feedback and positive prompt injection. Under the coverage information guidance, incremental iteration of test cases can extensively cover areas that were missed by the original test cases. In addition, due to the inability to count the number of focal methods correctly tested by EvoSuite, we compare the coverage of the two methods on the passed focal methods. That is, the denominator of the coverage calculation is the total number of branches/lines of the focal methods correctly tested by TestART. From Fig.~\ref{RQ2-F}, we observe that TestART outperforms EvoSuite in three projects, while in the remaining two, it trailed by an average coverage rate of only 5\%. This performance is notable given that EvoSuite, as an SBST method, seeks to cover source code by generating a large number of test cases. In contrast, TestART utilizes the powerful generative capabilities of LLMs to produce high-quality test cases in just half the quantity yet manages to achieve comparable quality to EvoSuite.

\begin{tcolorbox}[colback=gray!10, colframe=black, title=Answer to RQ2, boxrule=0.5mm, left=0mm, right=0mm, top=0mm, bottom=0mm]
\textbf{The test cases generated by TestART are much more sufficient to test the source code, proving the highest coverage rate compared to other baselines. TestART also achieves better testing sufficiency than EvoSuite with fewer test cases.}
\end{tcolorbox}

\subsection{Answer to RQ3 (Ablation)}

\textbf{Design.}
To validate the effectiveness of the different modules in TestART, we construct an ablation study of different parts of TestART to research its inner function. We conduct ablation experiments using three metrics on three incomplete and one complete TestART. 

\textbf{Results.} 
Table~\ref{rq3} shows the ablation study results of TestART, which demonstrates that the complete TestART achieves the best performance. We present the results for four different mutations of TestART: \textbf{Only ChatGPT-3.5, adding repair (+ Repair), adding repair and iteration includes positive prompt injection (+ Repair + Iteration), and adding repair, iteration, and coverage-guided testing feedback (Complete TestART).} From Table~\ref{rq3}, the repair strategy plays a vital role, increasing the Pass by 28.64\%, TBC by 19.03\% and TLC by 19.66\%. The iteration and coverage-guided testing feedback both contribute to the improvement of the coverage rate by about 3\%. 

\textbf{Analysis.} 
The results from Table~\ref{rq3}. show that different parts of TestART contribute to improving the quality of test cases. The most apparent core module for improvement is the repair strategy, which not only enhances the pass rate but also improves the coverage rate. This validates that our fixed templates do not improve pass rate at the expense of coverage. As mentioned in the motivation section, LLMs can generate high-quality test cases, but low pass rates limit the overall coverage rate. TestART solves the problem effectively through the empirical repair templates. In addition, positive prompt injection as a main part of iteration ensures that LLMs converge incrementally during loops. It is essential to suppress hallucination for LLMs; otherwise, the generated test cases will lead to a decrease in the pass rate.

\begin{tcolorbox}[colback=gray!10, colframe=black, title=Answer to RQ3, boxrule=0.5mm, left=0mm, right=0mm, top=0mm, bottom=0mm]
\textbf{The outcomes of the ablation study demonstrate that every component of TestART plays a crucial role in enhancing the quality of the generated test cases. The complete TestART achieves the highest coverage and pass rates.}
\end{tcolorbox}

\subsection{Answer to RQ4 (Generalization)}

\begin{table*}
\centering
\small
\caption{The experimental results of TestART and two baseline methods on HITS dataset.}
\label{RQ4}
\resizebox{\textwidth}{!}{
\begin{tabular}{p{0.8cm} p{1.7cm} p{0.7cm} p{0.7cm} p{0.7cm} p{0.7cm} p{0.7cm} p{0.7cm} p{0.7cm} p{0.6cm}|p{0.7cm} p{0.8cm}|p{0.6cm}} 
\toprule
Project & Method & CE$\downarrow$ & RE$\downarrow$  & Pass$\uparrow$  & TBC$\uparrow$  & TLC$\uparrow$  & BCCT$\uparrow$  & LCCT$\uparrow$ & TCC  & MC$\uparrow$  & TS$\uparrow$  & AC \\ 
\midrule
\multirow{3}{*}{\textbf{RUL}} & EvoSuite    & -       & -       &  57.15\% & 27.20\% & 38.64\% &  67.81\% &  84.63\% & 421& 37.40\% & 65.28\%  & - \\
&ChatGPT-3.5 & 47.88\% & 20.00\% & 32.12\% & 29.94\% & 30.98\% & 93.23\% & 96.45\% & 249& 32.36\% & 71.76\%  &181\\
&TestART     & 30.30\% & 8.48\% & 59.39\% & 58.59\% & 58.14\% & 98.63\% & 98.76\% & 300& 49.60\% & 75.40\% &320\\ 
\midrule
\multirow{3}{*}{\textbf{WIN}} &EvoSuite & -       & -       & 12.50\%  & 17.39\% & 11.33\% & 73.61\% & 74.12\% &  8& 11.47\% & 65.79\%  & - \\
&ChatGPT-3.5 & 50.64\% & 20.51\% & 27.56\% & 26.60\% & 26.42\% & 96.51\% & 95.57\% & 200& 38.07\% & 74.77\%  &92\\
&TestART     & 30.77\% & 7.05\%  & 60.26\% & 59.46\% & 58.54\% & 98.68\% & 96.19\% & 234& 59.17\% & 85.43\%  &260\\ 
\midrule
\multirow{3}{*}{\textbf{BPG}} &EvoSuite  & -       & -       & -       & -       & -       & -       & -       & -       & -   & -    & - \\
&ChatGPT-3.5 & 25.89\% & 10.11\% & 63.48\% & 62.78\% & 63.05\% & 98.72\% & 98.27\% & 655& 38.27\% & 87.35\%  &858\\
&TestART     & 18.62\% & 2.84\%  & 77.48\% & 75.85\% & 75.89\% & 98.74\% & 98.83\% & 7109&49.29\% & 88.68\%   &1120\\ 
\midrule
\multirow{3}{*}{\textbf{Total}} &EvoSuite   & -       & -       &  35.46\% & 22.43\% & 25.37\% & 70.63\% & 79.52\% & - &27.90\% & 65.35\% & - \\
&ChatGPT-3.5 &34.35\%  & 13.79\% & 51.30\% & 50.28\% & 50.61\% & 97.43\% & 98.09\% &1104 & 34.45\% & 56.94\%  &1131\\
&TestART     &\textbf{22.94\%}  & \textbf{4.63\%}  & \textbf{71.30\%} & \textbf{69.74\%}& \textbf{69.52\% }& \textbf{98.17\%} & \textbf{98.24\%} & 1253 &\textbf{53.11\%} & \textbf{79.20\%}  &\textbf{1700}\\
\bottomrule
\end{tabular}
}
\end{table*}

\textbf{Design.}
We set RQ4 primarily to verify TestART's performance on unlearned new data to avoid data leakage. Since HITS~\cite{wang2024hits} only experiments with certain complex focal functions within the dataset and does not validate all public focal methods of the entire project, we do not use this method as a baseline. We don't report EvoSuite's results on the BPG project because BPG is not executable on Java 1.8. We present the results of TestART along with two baseline methods shown in Table~\ref{RQ4}. The experiment results primarily include evaluation across three dimensions: correctness (CE, RE and Pass), coverage (TBC, TLC, BCCT and LCCT) and test case count.

\textbf{Results.} TestART clearly achieves superior performance in each project. In terms of overall HITS dataset comparisons, TestART increases correctness by approximately 20\% and surpasses ChatGPT-3.5 by 19\% in both test branch coverage (TBC) and test line coverage (TLC). Additionally, TestART generates 1,253 test cases for 885 focal methods, only 13\% more than ChatGPT-3.5.

\textbf{Analysis.}
Table~\ref{RQ4} clearly demonstrates that TestART maintains a high level of accuracy improvement even when facing untrained new data. The 20\% increase in Pass indicates that our proposed repair strategy is equally effective on new data. We also analyze the repair results, finding that \textbf{TestART successfully repaired 41.08\% of compilation errors and 71.70\% of runtime errors.} This aligns with the repair effectiveness observed on the Defect4J dataset. In addition, The nearly 20\% increase in both TBC and TLC further demonstrates that TestART achieves high testing effectiveness without significantly increasing the number of test cases. We also find that the correctness and coverage results on Defects4J and HITS dataset are almost indistinguishable, which suggests that the issue of data leakage does not significantly affect our method.

\begin{tcolorbox}[colback=gray!10, colframe=black, title=Answer to RQ4, boxrule=0.5mm, left=0mm, right=0mm, top=0mm, bottom=0mm]
\textbf{TestART demonstrates strong performance on the unlearned dataset, achieving a stable accuracy improvement while effectively generalizing to new data.}
\end{tcolorbox}

\begin{figure}[!tbp]
  \centering
  \subfloat[Time cost of TestART for single focal method.\label{fig:time}]{
    \includegraphics[width=0.38\textwidth]{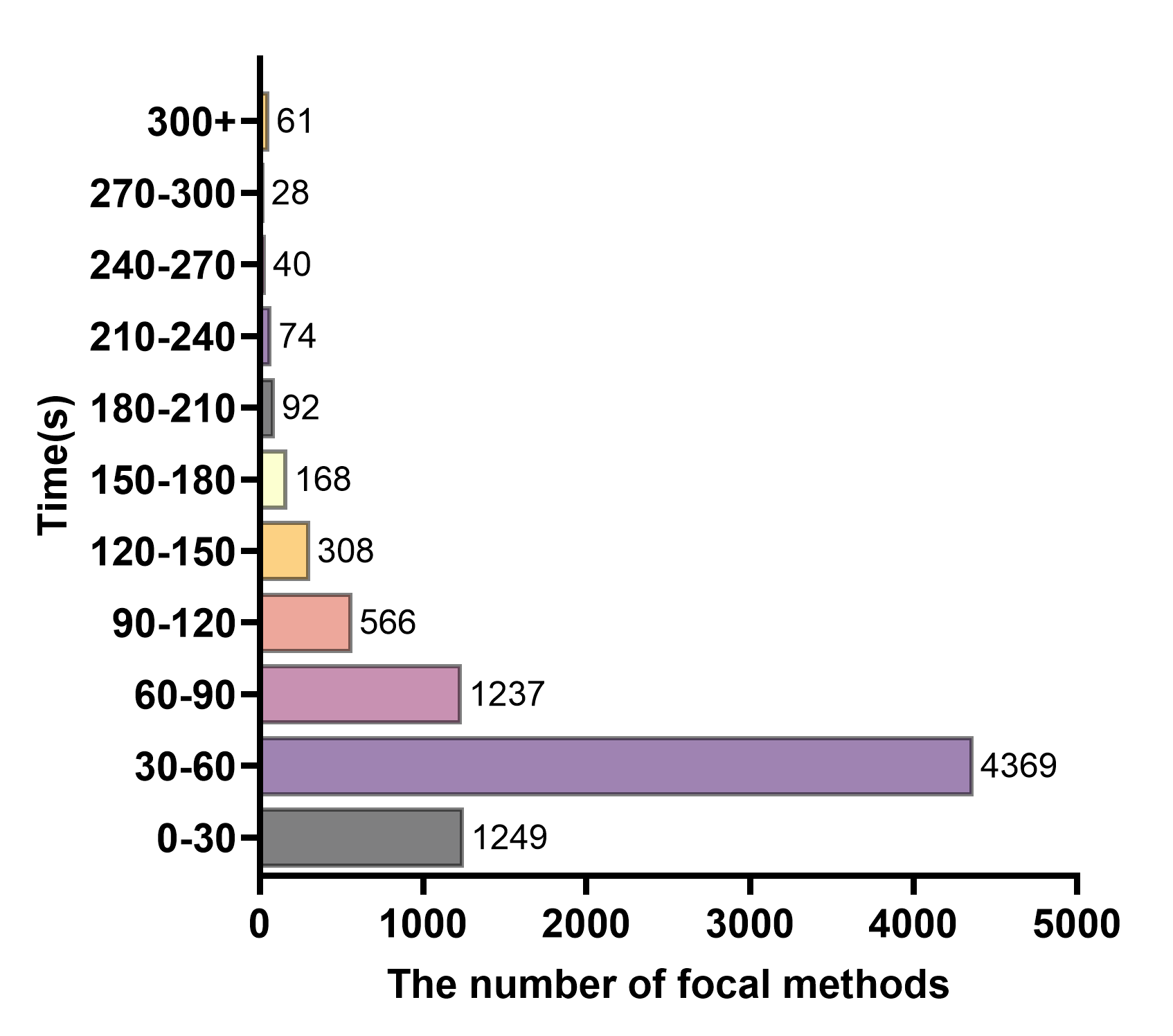}
  }
  \hfill
  \subfloat[Money cost of TestART for 100 generations of each project.\label{fig:money}]{
    \includegraphics[width=0.58\textwidth]{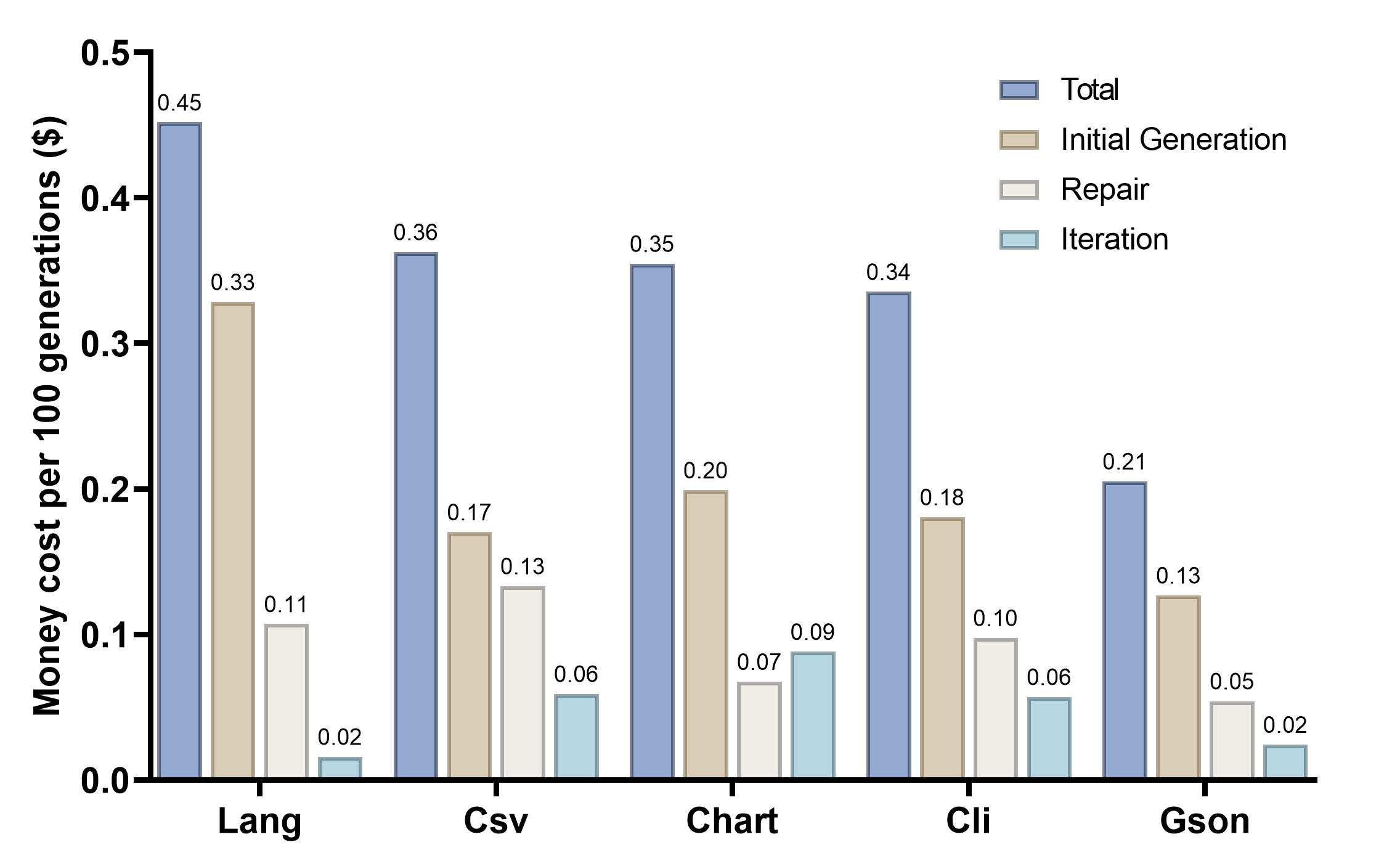}
  }
  \caption{Cost of TestART.\label{fig:cost}}
\end{figure}

\subsection{Answer to RQ5 (Cost)}
\textbf{Design.}
In software engineering, both time cost and monetary cost are crucial metrics for evaluating unit test generation tools. In terms of time cost, if the generation process takes too long, it will negatively impact development efficiency and reduce the tool's practicality. As for monetary cost, TestART utilizes ChatGPT by calling APIs, and each invocation incurs a certain fee. Whether this cost is reasonable is also a significant factor affecting the practicality of the unit test generation tool. This experiment evaluates the temporal efficiency of TestART by measuring the time required to generate unit test cases for individual focal methods. Additionally, it quantifies the monetary expenditure across three key phases—initial generation, repair, and iteration—for each project within the Defects4J benchmark.

\textbf{Results and Analysis.}
Fig.~\ref{fig:time} illustrates the time consumption of TestART for unit test generation tasks. As shown in the figure, 68.58\% of the unit test generation tasks are completed within one minute, while 90.59\% are finished within two minutes. Although a small portion of tasks requiring repair and iteration may consume more time, 99\% of the tasks can be completed within five minutes. The maximum search time we set when generating test cases using EvoSuite is 10 minutes, which means TestART is more efficient than EvoSuite.

Fig.~\ref{fig:money} presents the average cost of TestART for every 100 generations across different projects. The experiment calculates the cost based on the pricing of GPT-3.5-turbo-0125: 0.5\$/Mtoken for prompt and 1.5\$/Mtoken for completion. As shown in the figure, the total cost varies among different projects. Gson has the lowest unit test generation cost, averaging approximately 0.2\$ for 100 times unit test generation, while Lang requires about 0.45\$. Furthermore, the initial generation phase accounts for the largest proportion of the total cost, constituting approximately 50\%. The combined costs of code repair and iterative feedback make up the remaining 50\%. In some projects, such as Lang, the repair cost is higher than the iteration cost. This is because most focal methods in Lang are relatively simple, and once the test cases pass execution, they can achieve full branch and statement coverage without requiring further iterations. In contrast, for the Chart project, the iteration cost exceeds the code repair cost. This is attributed to the higher proportion of complex and large focal methods in this project, which often require multiple iterations to achieve high coverage, thereby increasing the cost of iterative feedback.

\begin{tcolorbox}[colback=gray!10, colframe=black, title=Answer to RQ5, boxrule=0.5mm, left=0mm, right=0mm, top=0mm, bottom=0mm]
\textbf{TestART demonstrates acceptable and controllable time and money cost, with 90.59\% of a single generation task completed within two minutes and the money cost for 100 generation tasks ranging from 0.3\$-0.45\$}
\end{tcolorbox}

\begin{figure*}[!tbp]
  \centering
  \includegraphics[width=0.9\linewidth]{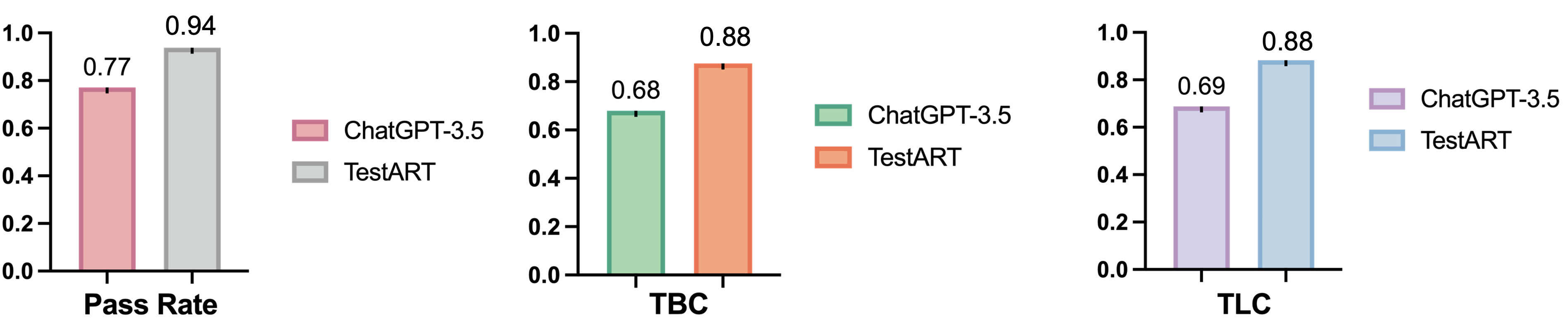}
  \caption{The experimental results of TestART on Internal dataset}
  \label{discussion_figure}
\end{figure*}

\section{Discussion}
\begin{itemize}[leftmargin=1em]
\item\textbf{Potential of error detection capability.} Error detection is one of the vital abilities of unit test generation. Therefore, we adopt mutation testing to evaluate the mutation-killing ability of TestART. As shown in the MC and TS columns of Table~\ref{RQ4}, TestART achieves a total of 53.11\% mutation coverage (MC) and 79.20\% test strength (TS). TestART outperforms EvoSuite by 20\% in the error detection scene. This result demonstrates that TestART can effectively identify mutation errors.

\item\textbf{Quality of assertion generation.} The number of assertions in a test case is important for measuring the quality of the test case. As shown in the last column of Table~\ref{RQ4}, the total number of assertions generated by TestART exceeds that of ChatGPT-3.5 by 569. This demonstrates that TestART not only improves the code coverage of test cases but also maintains a substantial number of assertions, ensuring comprehensive test validation.

\item\textbf{Applied to real-world industrial applications.} Unit test generation is widely used in the industry, making it crucial to conduct evaluations using real-world industrial data. TestART has already been integrated into Huawei's programming plugins based on the internal model to generate high-quality test cases. To further validate the effectiveness of TestART, we evaluate it on the real-world industrial dataset\footnote{Due to the confidential policy of the company, we used a closed-source model (e.g., ChatGPT-3.5) as a substitute for the internal model to conduct evaluation, and hide the name of internal projects.}. We present the results detailed in Fig.~\ref{discussion_figure}. This figure illustrates Pass, TBL and TLC on the Internal dataset. Notably, \textbf{TestART outperforms ChatGPT-3.5 across all three metrics, with significant improvements of about 20\% on Internal dataset}. This demonstrates that TestART is effective for both open-source datasets like Defects4J and the industrial dataset. 

\end{itemize}

\section{Threats to Validity}
\textit{\textbf{Baselines for comparison.}} The selection of baselines is one of the vital elements that threaten validity. Because the essence of unit testing is code generation, this type of method keeps emerging with diverse mechanisms. To alleviate this threat, we choose the state-of-the-art methods of four kinds of solutions as baselines, including the SBST tool, DL-based method, LLMs, and LLM-based approaches as the baselines, each empowered by different core engines.

\textit{\textbf{Dataset Selection.}} The selection of the source code dataset is another threat. Due to the large and varied datasets of unit tests, we chose three kinds of datasets (open-source, unlearned and industrial) to mitigate the threat of this choice and ensure fairness. The unlearned and industrial datasets are specifically employed to prevent data leakage from impacting the evaluation process.

\textit{\textbf{LLMs Selection.}} The last threat is the choice of the core LLMs. Our TestART can be applied to any interactive LLMs. We chose the most commonly and widely used LLM, ChatGPT-3.5, as the core engine. Although ChatGPT-4 outperforms ChatGPT-3.5, TestART (based on ChatGPT-3.5) achieves better results than ChatGPT-4. TestART has also been validated on other models, but we have not made the results publicly available for commercial reasons

\section{Related Work}

This section covers the work related to our proposal. We mainly introduce the automated unit test generation approaches and automated program repair techniques.

\subsection{Automated Unit Test Generation}
Automated unit test generation significantly improves test efficiency. SBST~\cite{mcminn2004search} uses metaheuristic search techniques (such as genetic algorithms) to automate or partially automate various testing tasks. The core of this approach lies in defining a fitness function for the specific testing problem, which guides the search algorithm to find effective solutions within a potentially infinite search space. AthenaTest~\cite{AthenaTest} uses a sequence-to-sequence transformer model to generate realistic, accurate and human-readable unit test cases. A3Test~\cite{A3test} further enhances the passing rate of AthenaTest by using a PLBART model verifying naming consistency and ensuring that test signatures match assertion knowledge. However, they suffer from low pass rates and depend excessively on fine-tuned datasets. 

In recent years, LLMs have performed excellently on unit test generation. In the literature~\cite{XiaoGLC24, abs-2406-18181,10329992,tang2024chatgpt,abs-2409-09464,abs-2305-00418}, Researchers empirically investigate the characteristics of unit test cases generated by large models and compare them with SBST tools, exploring the potential for combining the two approaches. Most methods are based on fine-tuning LLMs~\cite{DakhelNMKD24, abs-2406-15743,abs-2401-06765, PleinOKB24, shin2023domain, CodaMosa} based on the code LLMs~\cite{le2022coderl, wang2021codet5} or are designed through prompt engineering~\cite{wang2024hits, abs-2404-04966, abs-2403-16218, abs-2402-11910, alshahwan2024automated, Ryan0S00RR24, 3691620.3695513, abs-2310-00483, yuan2023no, schafer2023adaptive, xie2023chatunitest}. As a representative work in the field, TESTPILOT~\cite{schafer2023adaptive} automatically generates unit test cases for JavaScript programs without additional training or few-shot learning. The CODAMOSA~\cite{CodaMosa} proposed by Microsoft pioneers the combination of LLMs with SBST methods.  ChatUnitTest~\cite{xie2023chatunitest} develop the Generation-Validation-Repair framework to create an adaptive focal context that is integrated into prompts and then submitted to ChatGPT. ChatTester~\cite{yuan2023no} first understands the purpose of the focal method and then creates a unit test for it based on the help of the iterative step of intention generation. HITS~\cite{wang2024hits} proposes decomposing the method-to-test into slices and generating unit tests slice by slice, applying the ``divide-and-conquer'' algorithm. TestART demonstrates distinct advantages over ChatUnitTest~\cite{xie2023chatunitest} and ChatTester~\cite{yuan2023no} in both repair strategies and the iterative process. First, TestART's repair mechanism employs carefully designed templates derived from automated program repair (APR) research, enabling efficient automated repair of typical compilation and runtime errors in LLM-generated test cases. In contrast, ChatTester relies solely on unstable LLM-based error correction, while ChatUnitTest employs only basic syntactic and import repair rules. Second, TestART implements a coverage-driven iterative process that incrementally generates test cases through the co-evolution of generation and repair. This differs fundamentally from ChatUnitTest's non-iterative approach and ChatTester's feedback mechanism neglects coverage-guided information.

In addition, researchers focus on the understandability~\cite{abs-2408-11710} by contextualizing test data, enhancing identifier names, and adding descriptive comments or measuring the readability~\cite{abs-2407-21369} of LLM-generated unit test cases. Shin et al.~\cite{abs-2409-12682} investigate the efficacy of RAG-based LLMs in test generation. In addition to mainstream research on Java unit testing, there are also approaches for other languages. CLAP~\cite{abs-2407-21429} is a Chat-like execution-based assert prediction for generating meaningful assert statements for Python projects. Karanjai et al.~\cite{abs-2407-05202} explore the capabilities of generative models in crafting unit testing cases for parallel and high-performance software with C++ parallel programs. To better evaluate the unit test generation methods, scholars have proposed benchmarks~\cite{zhang2024testbench, abs-2408-07846} to measure the capability of unit test generation. TestSpark~\cite{SapozhnikovOPKD24} is a plugin for IntelliJ IDEA that enables users to generate unit tests with only a few clicks directly within their Integrated Development Environment (IDE). Among this, some researchers have also paid attention to test case repair~\cite{abs-2407-03625, abs-2401-06765, schafer2023adaptive}.

These methods often achieve reasonable coverage rates and generate test cases with high code readability. Nevertheless, LLMs may face limitations such as getting stuck in compile and runtime errors more frequently. Unlike previous works focusing more on the generation process, TestART applies the APR technique to repair the errors in the generated test cases rather than regeneration or direct LLM-based repair. TestART aims to improve pass and coverage rates by using repair templates and coverage-guided testing feedback.

\subsection{Automated Program Repair}
In the literature, APR techniques are mainly categorized into four groups~\cite{zhang2023survey,gazzola2019automatic,monperrus2018automatic}, i.e., heuristic-based~\cite{le2012genprog}, constraint-based~\cite{xiong2017precise}, template-based~\cite{liu2019tbar}, and learning-based~\cite{yuan2022circle} repair techniques. Our work is related to template-based APR, which is discussed below.

Template-based APR attempts to directly transform the buggy code into the correct one based on pre-defined fix patterns and represents state-of-the-art~\cite{kim2013automatic,liu2019tbar,koyuncu2020fixminer,zhang2023gamma}.
As the flagship work in this field, TBar~\cite{liu2019tbar} systematically summarizes existing fix patterns and applies these patterns to patch generation. Besides, FixMiner~\cite{koyuncu2020fixminer} leverages a three-fold clustering strategy to extract fix patterns based on AST represent, and AVATAR~\cite{liu2019avatar} exploits fix pattern of static analysis tools to generate patches. Recently, inspired by the potential of combing fix patterns and LLMs, GAMMA~\cite{zhang2023gamma} frames APR as a fill-in-the-blank task by querying LLMs to directly recover the correct code for masked cod with the code context. 
According to ReAssert~\cite{daniel2009reassert}, repairing unit tests by empirical strategies is effective and acceptable for developers. However, this strategy is not for test cases generated from LLMs.

Therefore, to increase test case passing rates, TestART designs a template-driven repair strategy that is adaptive to LLM outputs inspired by the APR community. Unlike template-based APR techniques, which usually focus on semantic bugs from production code, TestART attempts to design simple and efficient templates to repair the typical compilation and runtime errors from test cases automatically generated by LLMs. More importantly, TestART implements the co-evolution between the generation and repair process to generate high-quality test cases under several iterations.

\section{Conclusion}
This paper presents TestART, the first approach to integrate the traditional automated repair technique with the generative capabilities of LLMs through an innovative co-evolutionary framework for generating high-quality unit test cases. TestART also introduces positive prompt injection and coverage-guided testing feedback to mitigate the effects of faithfulness hallucinations in LLMs and enhance the sufficiency of test cases. TestART significantly outperforms existing methods, showing an 18\% increase in passing rate and a 20\% enhancement in coverage rate on tested methods, marking substantial improvements over the capabilities of previous works. Although TestART is experimented on the ChatGPT-3.5 model, it is superior to the ChatGPT-4.0 model and can be implemented in other LLMs. TestART shows excellent performance on both open-source datasets and industrial datasets. This indicates that TestART effectively leverages LLMs' strengths while mitigating their weaknesses, leading to more effective, reliable, and higher-quality unit tests.

\section{Acknowledgment}
This work is supported partially by the National Natural Science Foundation of China (61932012, 62372228), and CCF-Huawei Populus Grove Fund (CCF-HuaweiSE202304, CCF-HuaweiSY202306).

\section{Data Availability}
Our experimental materials are available at \url{https://github.com/sikygu/TestART}.

\bibliographystyle{ACM-Reference-Format}
\bibliography{reference}

\end{document}